\documentclass[lettersize,journal]{IEEEtran}
\usepackage{amsmath,amsfonts}
\usepackage[]{hyperref} 
\usepackage{xcolor}
\definecolor{citeblue}{RGB}{0,0,255} 
\hypersetup{
  colorlinks=true,
  linkcolor=citeblue,
  citecolor=citeblue,
  urlcolor=citeblue
}
\usepackage[linesnumbered,ruled]{algorithm2e}
\usepackage{array}
\usepackage[caption=false,font=normalsize,labelfont=sf,textfont=sf]{subfig}
\usepackage{textcomp}
\usepackage{stfloats}
\usepackage{url}
\usepackage{verbatim}
\usepackage{graphicx}
\usepackage[justification=centering]{caption}
\usepackage{cite}
\usepackage[normalem]{ulem}
\useunder{\uline}{\ul}{}

\begin{document}

\title{HyperFedNet: Communication-Efficient Personalized Federated Learning Via Hypernetwork}

\author{Xingyun Chen, Yan Huang,  Zhenzhen Xie,  Junjie Pang

\thanks{
Xingyun Chen and Junjie Pang are with the College of Computer Science and Technology, Qingdao University, Qingdao, Shandong 266000, China. (e-mail: 2021020697@qdu.edu.cn; pangjj@qdu.edu.cn)

Yan Huang is with the College of Computing and Software Engineering, Kennesaw State University, Atlanta, GA, USA. (e-mail: yhuang24@kennesaw.edu)

Zhenzhen Xie is with the School of Computer Science and Technology, Shandong University, Qingdao,Shandong 266237, China. (e-mail: xiezz21@sdu.edu.cn)}
\thanks{Junjie Pang is the corresponding author.}
}

\markboth{Journal of \LaTeX\ Class Files,~Vol.~14, No.~8, August~2021}%
{Shell \MakeLowercase{\textit{et al.}}: A Sample Article Using IEEEtran.cls for IEEE Journals}

\IEEEpubid{}

\maketitle

\begin{abstract}
In response to the challenges posed by non-independent and identically distributed (non-IID) data and the escalating threat of privacy attacks in Federated Learning (FL), we introduce HyperFedNet (HFN), a novel architecture that incorporates hypernetworks to revolutionize parameter aggregation and transmission in FL. Traditional FL approaches, characterized by the transmission of extensive parameters, not only incur significant communication overhead but also present vulnerabilities to privacy breaches through gradient analysis. HFN addresses these issues by transmitting a concise set of hypernetwork parameters, thereby reducing communication costs and enhancing privacy protection. Upon deployment, the HFN algorithm enables the dynamic generation of parameters for the basic layer of the FL main network, utilizing local database features quantified by embedding vectors as input. Through extensive experimentation, HFN demonstrates superior performance in reducing communication overhead and improving model accuracy compared to conventional FL methods. By integrating the HFN algorithm into the FL framework, HFN offers a solution to the challenges of non-IID data and privacy threats.
\end{abstract}


\section{Introduction}
\IEEEPARstart{I}{n} recent years, there has been a growing emphasis on data security due to increased awareness of privacy protection and the evolution of laws. However, in the era of big data, the continuous generation of diverse data by electronic devices poses challenges. This data contains personal privacy information and is unsuitable for transmission to centralized servers for data mining or machine learning purposes. For example, input prediction technology has significantly enhanced the accuracy of word suggestions, making typing more convenient and enjoyable. While training neural networks requires substantial data support, many owners of private data are reluctant to share it. Conversely, there is an increasing availability of computing resources from various devices. In this context, Federated Learning (FL)\cite{fedavg-mcmahan2017communication} has emerged as a solution to address this dilemma. FL is a novel distributed technique that enables multiple parties to collaborate in training neural network models by transmitting only the parameters of the training network without sharing the dataset. This collaborative approach allows users to achieve common goals and benefit from the learned models while safeguarding their privacy. However, the development of FL faces challenges related to communication cost and data heterogeneity. As device computing power increases, the complexity of model structures grows, necessitating the transmission of hundreds of thousands or even millions of parameters between users' devices and the server per round. Consequently, the communication traffic overhead for users becomes substantial. This issue often leads to devices participating only when connected to WiFi, limiting the number of available devices and resulting in suboptimal model learning for FL tasks. Additionally, data heterogeneity arises from variations in user behavior, domain differences, and the disparity in data volume between individual users and institutional users. Data heterogeneity poses a significant obstacle that hinders the improvement of model accuracy and can even impede model convergence. Effectively addressing communication resource conservation and personalized federated learning (pFL) are crucial yet challenging aspects in the field of FL. Existing methods in FL have not adequately tackled these two challenges simultaneously.

Our work introduces a novel algorithm called HyperFedNet (HFN) to address the challenges of personalized training tasks and heterogeneity in FL. HFN leverages the advantages of Hypernetworks\cite{ha2016hypernetworks} to transmit a small number of parameters for personalized training in FL. Hypernetworks are neural networks that generate parameters for another network, allowing for more efficient communication. To tackle the heterogeneity problem in FL, we propose a representation method that quantifies different heterogeneous features using embedding vectors. These embedding vectors serve as representation vectors, capturing the characteristics and diversity among different objects. By mapping objects to different locations in the embedding vector space, we can effectively address data heterogeneity. In FL, users can utilize features learned by other users to improve their own model performance. The communication in FL aims to enable the main network to learn from users' strengths. By leveraging the generalization ability of neural networks and using a small neural network to generate appropriate parameters for the main network based on different embedding vectors, we can effectively handle traffic overhead. Instead of communicating with the main network, we design a hypernetwork that communicates parameters via input embedding vectors. Each local model is represented by a unique and learnable embedding vector, serving as input to the hypernetwork. This allows the hypernetwork to specify the characteristics of the local model structure and provide appropriate outputs. The hypernetwork acts as a coordinate map of a low-dimensional manifold in the embedding vector space, where each unique user model structural parameter is constrained and parameterized by the embedding vector\cite{fedpFedHN-shamsian2021personalized}.

Our contribution are listed as follows:
\begin{itemize}
\item We propose a pFL method that reduces communication overhead and improves accuracy compared to existing algorithms, offering a tailored balance between model accuracy and communication needs.

\item We introduce an innovative parameter transmission strategy in FL, utilizing aggregation of small model parameters, which reduces transmission costs and improves data security.

\item Our extensive experiments demonstrate the proposed HFN method's superiority over current FL algorithms in communication efficiency, convergence speed, and personalization accuracy.

\item We highlight the integration compatibility of the HFN algorithm with existing algorithms, showing improved performance and potential to enhance current FL approaches.
\end{itemize}

In summary, our work contributes by proposing a pFL method that reduces communication volume and improves security through a novel parameter transmission idea. Through extensive experiments, the paper demonstrates the effectiveness of the proposed method and its potential for integration with other algorithms.

The paper is organized as follows: Section II presents some work done by other researchers in related areas. Section III adds some necessary background knowledge and problem definition. In Section IV, the HFN method is formulated. The experiments in Section V show the advancement of HFN and the effect of combining it with other algorithms. In Section VI we summarize our work.

\section{Related Work}
\subsection{Personalized Federated Learning}
The pFL is a way to solve the dilemma of FL in the face of data heterogeneity. Users' and organizations' location, country, habits, or research directions lead to wide variations in the type and amount of local data stored on the device\cite{tan2022towards}. For example, city cameras are more likely to capture pictures of humans and cars compared to forest cameras. Many researchers have worked on solving the problems posed by this data heterogeneity. 

Researchers have proposed many pFL methods. Data augmentation is a technique that transforms or expands the original data to generate new training samples to balance data heterogeneity. In order to balance the differences between clients, some researchers have proposed the use of data augmentation\cite{duan2020self,zhao2018federated,jeong2018communication}. Due to the feature detection limitations of local models, some algorithms try to add proximal terms\cite{fedProx-li2020federated,fedDitto-li2021ditto,fedDyn-acar2021federated,fedSCAFFOLD-karimireddy2020scaffold}. The training process of the local model is constrained and adapted to force a closer approximation of the global model's objective so that a more suitable model can be obtained. This approach can be achieved by adding appropriate penalty or regularization terms to the objective function. In the case of large differences in the probability of occurrence of different categories in the database, the model may favor the more frequently occurring categories and perform poorly on the less frequent ones, which may lead to a decrease in the overall accuracy of the model. \cite{fedlc-zhang2022federated} is based on the probability of occurrence of each category, and logits are calibrated before softmax cross-entropy in the expectation of higher accuracy. A single global model may not fit well when faced with non-iid user data. Partially pFL learns a unique model for each user to get better learning gains. These algorithms try to go for the balance between shared knowledge and personalization. \cite{fedPer-arivazhagan2019federated} aggregates only the basic layer while retaining the personalization layer, but requires all users to participate in every round, which is not feasible in the FL environment. \cite{fedMD-edli2019fedmd,fedPerFedAvg-fallah2020personalized} finds an initial shared model based on migration learning, knowledge distillation, and meta-learning, respectively, and based on the initial model, the users fine-tune the model locally to obtain their own personalized model. In addition, \cite{fedBabu-oh2021fedbabu} fixes the personalization layer during training, and after convergence, the local data is used to fine-tune the model to obtain a localized model.

The use of data augmentation, meta-learning, and other techniques in FL introduces the risk of data leakage, as it involves some form of data sharing, which contradicts the original purpose of FL. In contrast, the HFN adheres to the fundamental principles of FL without exchanging any data. While adding proximal terms to the model is a simple approach, it carries the risk of overfitting and reduces the model's generalization ability. This approach relies on the existence of a global model as a reference to guide the local models towards convergence. However, the quality and accuracy of the global model are crucial for the effectiveness of pFL. If the global model is flawed or biased, it can negatively impact the training of personalized models. In the HFN algorithm, the model's effectiveness does not solely rely on the global model but also incorporates the embedding vector and personalization layer, which significantly reduces the risk of overfitting. Calibration methods in FL usually make assumptions about the fixed and known data distribution, assuming that the probability of each category's occurrence remains constant. However, real-world data distributions are often dynamic, and the probabilities of category occurrences may change over time or in different environments. This can lead to calibration method failures or the need for frequent adjustments. In contrast, the HFN implicitly represents the probabilistic features of the user database using a embedding vector and a personalization layer. These dynamic parameters continuously update and adapt as the model learns. A single global model is not well-suited for highly non-independent and homogeneously distributed data scenarios. Training the network using the model decoupling approach still subjectively transmits all feature knowledge to the user without targeting personalized knowledge for the main network. As a result, users are unable to obtain their own unique knowledge, leading to slow convergence, high communication overhead, and poor modeling results. The HFN algorithm, on the other hand, generates different knowledge for different users, enhancing convergence speed, and improving the model's effectiveness.

\subsection{Efficient Communication in Federated Learning}
FL is a distributed machine learning approach in which model parameters and updates need to be exchanged frequently between users for collaborative model training. The large number of parameter transfers leads to expensive communication overheads for the users. The high communication cost not only has an impact on the overall performance, but also poses a challenge to devices with limited bandwidth and energy, thus constraining the development of FL\cite{li2020federated1}. As a result, several research efforts have been devoted to reducing the amount of communication between users. A common approach is to compress and encode model parameters using compression techniques to reduce the bandwidth required for transmission\cite{shah2021model,haddadpour2021federated,stich2020communication}. For example, some researchers have utilized global models to construct submodels and send the submodels to the users through lossy compression. Each user performs local updates on the smaller model and provides updates that can be applied to the larger global model on the server. The whole process reduces the transmission by constructing the submodel once and compressing it twice \cite{caldas2018expanding}. On the other hand, some work focuses on reducing the frequency of communication. For example, researchers have proposed strategies to increase the number of local computation epochs\cite{fedavg-mcmahan2017communication}. However, determining the optimal number of local computation epochs is a key issue, and to address this issue. The study by Reisizadeh et al.\cite{reisizadeh2020fedpaq} introduces the periodic averaging method and quantizes its updates based on the quantizer's accuracy before uploading to the parameter server, thereby lowering communication overhead. In addition, several research works look at reducing the communication overhead in hierarchical federated learning systems\cite{he2023hierarchical}. For example, Deng et al.\cite{deng2021share} proposed an algorithm to minimize the communication overhead from edge aggregation and cloud aggregation. Some academics combine methods from other fields, He et al.\cite{he2020group} proposed to utilize knowledge migration instead of traditional transmission transfer, where instead of sending the network parameters back to the server, the user sends feature maps with fewer parameters and associated logs to the server, which utilizes this information to train a convolutional neural network with a larger model.

Compressing and encoding model parameters can introduce noise and lead to information distortion, potentially resulting in reduced model performance. Additionally, the compression and decompression processes between users and the server incur computational overhead. This overhead becomes particularly significant for large models, which may outweigh the benefits gained from compression. Moreover, when employing optimization methods that increase local computing rounds, the heterogeneity among users, including differences in computing capabilities and data distribution, can pose challenges. Adding more local computational epochs may further amplify this variability, leading to poorer training performance for certain users. Quantization operations, similar to model compression, can result in information loss. Representing the model with low-precision values may not accurately capture the intricacies and complexity of the model, necessitating more training iterations to compensate for the loss of detail. Moreover, when areas such as knowledge transfer are combined with FL, transmitting only feature maps and related logs without transmitting network parameters can lead to information loss. Consequently, the server can only use local information for training and may not be able to acquire a complete understanding of the model.

\subsection{Federated Learning with Hypernetworks}
The concept of fast weighting, in which one network can produce context-dependent weight changes for a second network, was introduced in \cite{schmidhuber1992learning}. With the advancement of hardware and the boom of machine learning in the past few years, many papers\cite{bertinetto2016learning,jaderberg2017decoupled,denil2013predicting,yang2015deep} have investigated this aspect. Similar hypernetworks have been widely used in various machine learning applications, such as neural architecture search\cite{brock2017smash}\cite{zhang2018graph}, computer vision\cite{Qiao_2018_CVPR}\cite{su2020blindly}\cite{alaluf2022hyperstyle}. Continuous Learning\cite{von2019continual}, FL\cite{fedpFedHN-shamsian2021personalized}\cite{fedpFedLA-ma2022layer}, Multi-objective Optimization\cite{navon2020learning}.

The first application of hypernetwork to FL is the pFedHN algorithm\cite{fedpFedHN-shamsian2021personalized}. It generates parameters for the basic layer of pFL by using a larger, more heavily parameterized, and computationally heavy hypernetwork on the server side than the main network. The authors show that FL architectures that incorporate hypernetworks provide cross-user parameter sharing while maintaining the ability to generate unique and diverse user models. \cite{fedpFedLA-ma2022layer} deploys a separate hypernetwork for each user on the server, which discerns the importance of each layer in the aggregation of models from different networks, and thus gives layer weights to the users under its charge to progressively exploit similarities between users and personalize models. In model framework heterogeneous federated learning, \cite{litany2022federated} utilizes hypernetworks to generate model parameters for different architectures. Hypernetwork has made a big splash in various aspects of FL, but its huge potential has not been fully exploited, and the task of how to use hypernetwork to achieve pFL for efficient learning with low communication is given to HFN.

In our work, we propose a novel approach that combines hypernetworks with FL. The HFN algorithm involves using a small hypernetwork that has fewer parameters and requires less computing power. By utilizing this hypernetwork, we significantly reduce the transmission cost. Rather than solely outputting hierarchical model weights, HFN's hypernetwork directly generates model parameters. This allows us to learn these parameters directly. By adopting this strategy, we aim to improve the efficiency and effectiveness of the FL training process.

\section{Preliminaries}
\subsection{Federated Learning}
Assume that in a FL scenario, there is a FL server and $K$ users, where user $k(k\in \{1,2,\dots,K\})$ has a personal data set $\mathcal{D}_k$. Before training, the FL server will initialize a global model to solve the objective function. In each iteration cycle, the server randomly selects some users to participate in training and delivers the global model parameters. After the selected user receives the global model, it is set as a local model and trained using its local data. The training process can use standard machine learning algorithms and optimization techniques, such as stochastic gradient descent. After training, the locally updated model parameters are transmitted back to the server. The server uses an aggregation algorithm to aggregate the parameters uploaded by the user, generates a new global model, and performs the next round of iterative training until the preset accuracy is reached or the model converges. During the FL training process, a set of parameters $w$ is found for the global model to minimize the overall loss. The objective function is defined as:

\begin{equation}
\operatorname*{min}_{w}\left[\frac{1}{K}\sum_{k=1}^{K}f_{k}(w)\right]
\end{equation}

Among them, $f_{k}$ represents the loss function of user $k$, which is defined as:
\begin{equation}
f_k\left(w_k\right)=\mathbb{E}_{\left(x,y\right)\sim p_k}\mathcal{L}\left(w_k;\left(x,y\right)\right)
\end{equation}
where $p_k$ represents the probability distribution of data set $\mathcal{D}_k$ generated by user $k$, and $\mathcal{L}\left(w_k;(x,y)\right)$ represents the prediction loss function.

Through FL, the entire process is learned without direct access to the original data, and the global model gradually incorporates the knowledge held by the participating user databases. Since data is kept locally and only model parameters are transmitted, FL provides a privacy-preserving mechanism that helps reduce the risk of data leakage and privacy violations. Especially in scenarios where sensitive data needs to be processed, such as healthcare, finance, and the Internet of Things. It enables participating parties to jointly train powerful machine learning models while ensuring data privacy and security.

\subsection{HyperNetworks}

\begin{figure}
\centering
\includegraphics[width=1.0\linewidth]{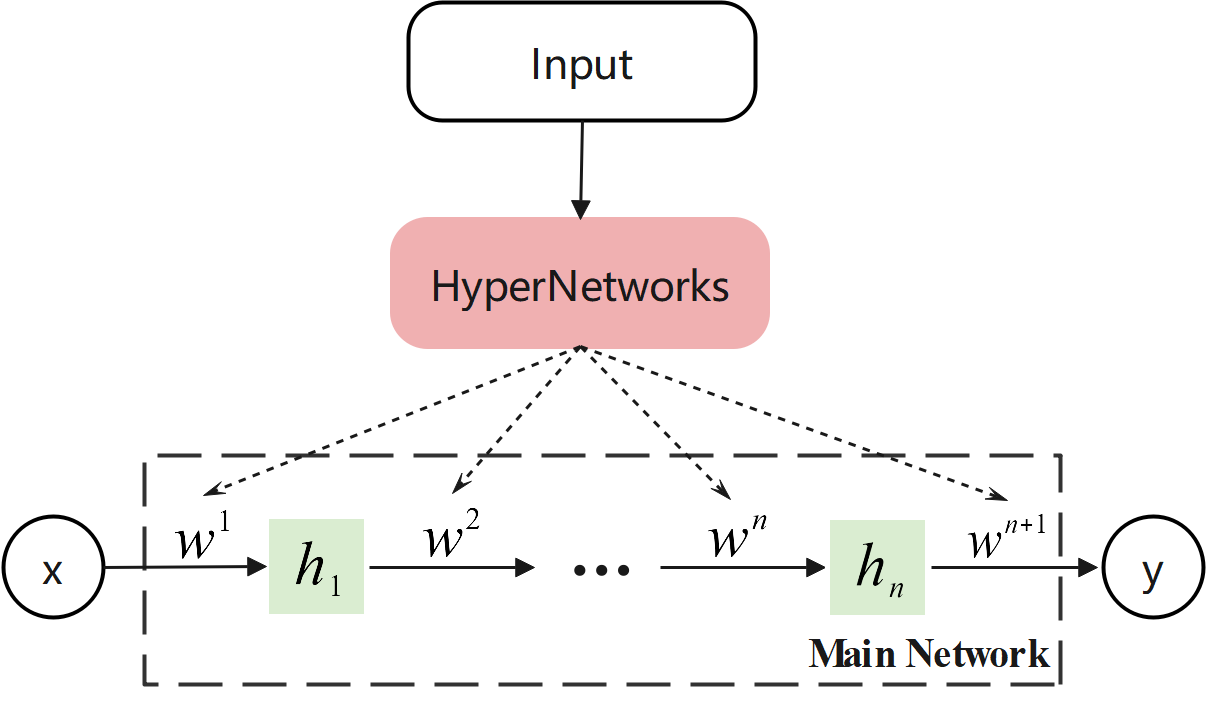}
\caption{Hypernetworks diagram}
\label{fig:Hypernetworks}
\end{figure}

Hypernetwork is a special type of neural network structure, usually containing one or more hidden and output layers, which has dynamically adjustable parameters. Hypernetworks can be used to generate weights, parameters, or architectures for other neural networks, thus enabling adaptive or automated network design during model training. As shown in Fig.\ref{fig:Hypernetworks}, a hypernetwork accepts a parameter sampling vector as input during training. The hypernetwork uses this input to generate parameters for other neural networks, which are generally called the main network. The parameters of the main network are no longer learned by the main network itself, but are updated by receiving the output of the hypernetwork. The hypernetwork transforms this input into parameters for other neural networks through a complex internal mechanism. These generated parameters can be seen as a kind of search space for exploring the possibilities of different network architectures or weights. The generated parameters are used to construct a specific neural network architecture. This process can be translated into an actual network architecture by some predefined rules or algorithms, e.g., using generative adversarial networks or by hypernetwork tuning. The generated network architecture is used for forward and backpropagation of the training data. Typically, the output is computed by forward propagation using the generated network weights, then the loss is computed based on a given loss function and the parameters of the hypernetwork are updated by a backpropagation algorithm. During training, optimization can be performed by iteratively updating the parameters of the hypernetwork and generating the sampling vectors of the network. This process can be implemented using traditional gradient descent or other optimization algorithms.

\subsection{Problem Definition}
In ordinary FL, all users share a global model and use their respective local data for training. However, different users may have different data distributions and characteristics. pFL allows each user to build his or her own personalized model to better adapt to the characteristics of local data. pFL can better improve the performance and adaptability of the model.

The personalization algorithm HFN we designed can generate $K$ sets of parameters based on its own characteristic input, and splice it with a locally retained personalization layer, so that each user can learn parameters $w_k$ that adapt to the local data distribution $p_k$ to achieve better forecasting. That is, the goal of HFN can be expressed as:
\begin{equation}
w^{*}=\arg\min_{w}\left[\frac{1}{K}\sum_{k=1}^{K}f_{k}(w_{k})\right]
\end{equation}
Where $w=\{w_1,...,w_K\}$ is the personalized parameter set of all users. $w^{*}$ is the current optimal solution obtained through continuous iterative learning as FL progresses.

In order to solve $w^{*}$ for personalization while achieving privacy enhancement and communication efficiency, our work expects to explore a new FL framework in which each user is able to complete the training by simply uploading a summary of the users' local model. Therefore, the pFL problem of our work turns to how to establish a way to automatically generate model summaries that adaptively complete the training of FL under external conditions.

\section{Methdology}
\subsection{Framework Overview}

\begin{figure*}
\centering
\includegraphics[width=0.75\linewidth]{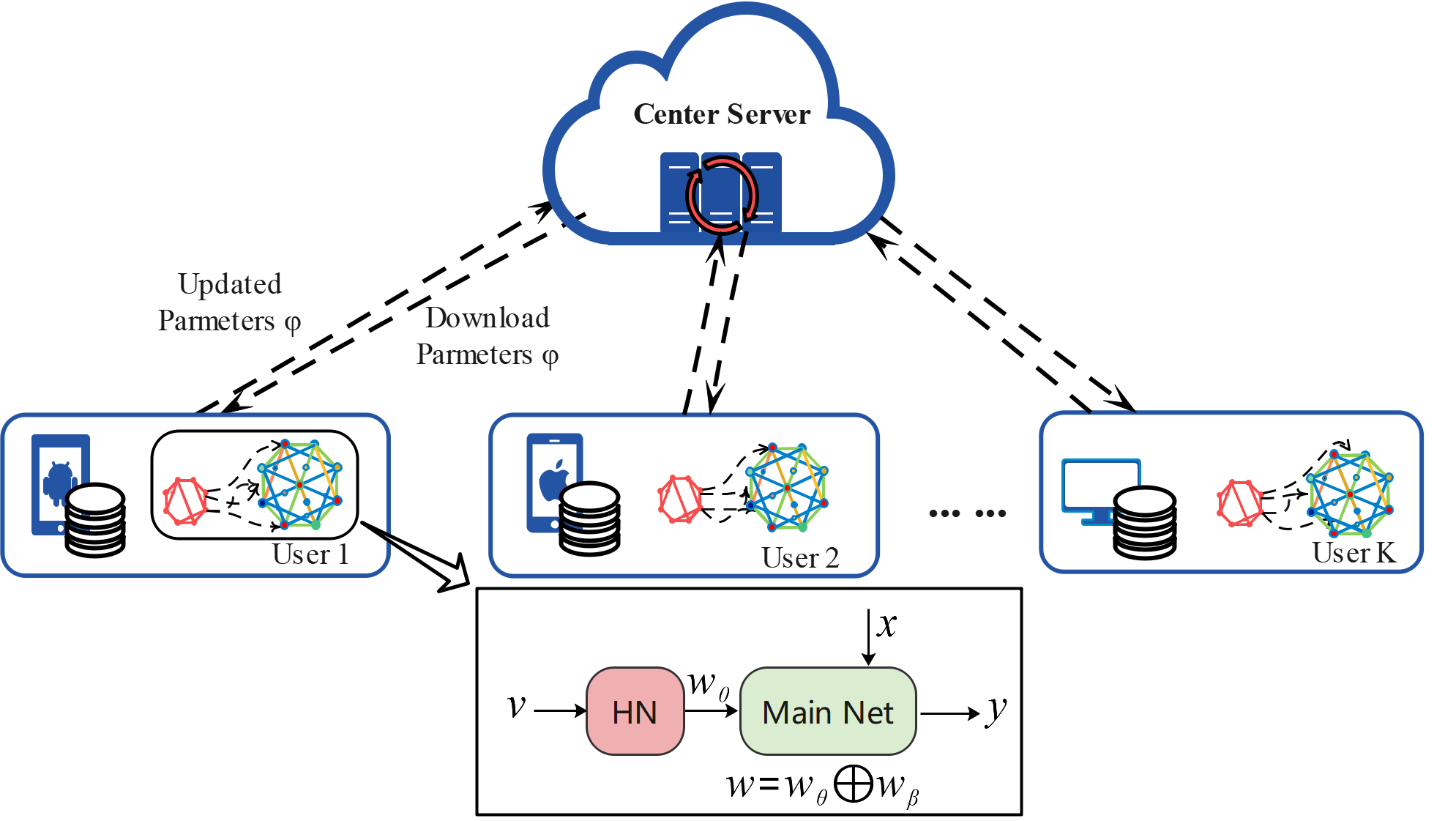}
\caption{HFN Framework: Small red networks are hypernetwork and large colored ones are main networks. During parameter aggregation, only the hypernetwork parameters are aggregated.}
\label{fig:HFN}
\end{figure*}

In FL, the process of training the global model through mutual learning between users is actually achieved through parameter exchange. More specifically, new global model parameters are generated based on a certain aggregation algorithm. The hypernetwork also has the ability to generate parameters, and can complete model generation for different users through adaptive external input. Therefore, we try to embed the hypernetwork into FL and let it complete the learning instead of the main network.

We propose the HFN algorithm with the architecture shown in Fig.\ref{fig:HFN}, the model running on the user consists of two parts, one for the hypernetwork and the other for the FL main network. The hypernetwork takes the user's local embedding vector as input and outputs parameters to the basic layer of the main network based on the characteristics of its representation. In the main network, it accepts the parameters and will perform the prediction training task locally. As input embedding vectors $v$ to the hypernetwork, they correspond to different filters, which are used to characterize some potential features of the filters, essentially reacting to the characteristics of different user databases, in order to help the hypernetwork to accomplish the output localization, as well as to better cope with non-iid data. These embedding vectors $v$ are fed as inputs into the hypernetwork HN to generate the basic layer parameters $w_{\theta}$ of the main network. In the setting, each user's classification layer is retained as a personalization layer in the respective main network model, and the retention of these layers contributes to personalization. The basic layer parameter $w_{\theta}$ is combined with the personalization layer parameter $w_{\beta}$ to consist of the complete main network parameter $w$. When performing forward propagation, the corresponding prediction $y_i$ is made based on the input $x_i$. If it is in the training phase, a local update is also required, and the update of the hypernetwork in the backpropagation is based on the gradient of the main network. In the parameter exchange phase, the HFN architecture communicates by transmitting the hypernetwork parameters $\varphi$. This small and sophisticated network, through aggregation, can communicate its own learned knowledge and ability to generate parameters with other users, and continuously improve the ability to refine the output main network parameters.

\subsection{pFL Based on Hypernetwork}

\begin{figure}
\centering
\includegraphics[width=1.0\linewidth]{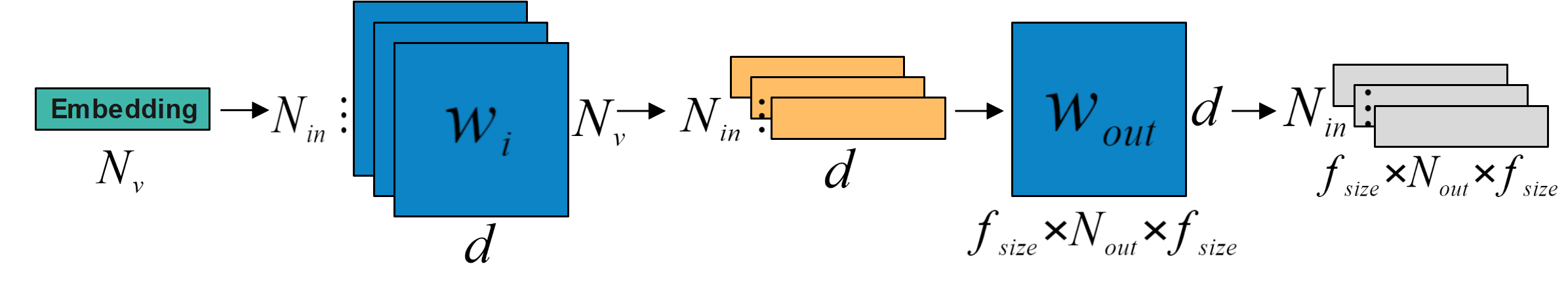}
\caption{Hypernetwork Model Structure.}
\label{fig:Model}
\end{figure}

Optimizing the communication overhead for FL is an important issue. Computer vision tasks occur frequently in FL. Convolutional Neural Networks (CNN) are the main algorithms for computer vision tasks. CNN extract abstract features of the input data layer by layer through multiple convolutional layers. In a typical CNN, most of the model parameters are located in the filters of the convolutional layers. However, traditional transmission methods directly use the CNN model as a global model, and directly transmitting the parameters of each filter in a CNN leads to heavy communication overhead. We attempt to adopt an approach that takes advantage of the powerful generalization and mapping capabilities of the hypernetwork to generate a large number of parameters in the convolutional layer of the main network by mapping them through the hypernetwork. First, the main network parameter $w$ is decomposed into the basic layer parameter, i.e., the convolutional layer parameter $w_{\theta}$, and the personalized layer parameter $w_{\beta}$. Then the process of communicating the aggregate of the basic layer parameter $w_{\theta}$ is transformed into a learning problem by introducing a hypernetwork, which allows the hypernetwork parameter $\varphi$ with fewer parameters to be aggregated instead of the main network, thus reducing the communication overhead in FL. At the same time, we keep the personalized layer parameter $w_{\beta}$ locally, and each user can personalize the training and adjustment according to his/her own data characteristics. This helps to improve the adaptability and performance of the model to better fit the characteristics of local data.

The convolutional layer in a convolutional network is composed of many filters. Every filter extracts a specific texture\cite{xia2016every}. Each filter for each user needs to have a embedding vector $v^{j}_{k}$ to represent the potential features of the filter. The purpose of quantizing the convolutional layer filters by embedding vector is to extract potential features. The embedding vector will help the hypernetwork to specify what kind of texture needs to be detected by the filter parameters that will be generated. The learnable embedding vector is denoted by $v^{j}_{k}$. Where $k$ denotes the user number and $j$ denotes the number of filters in the convolutional layer. The size of the embedding vector can be flexibly adjusted, A larger embedding vector can capture more feature information, and provide richer input information to the hypernetwork. However, it should be noted that increasing the size of the embedding vectors also leads to an increase in the number of parameters $\varphi$ of the hypernetwork, which increases the communication overhead, since these parameters need to be transmitted in the framework we designed. If the task requires more feature information for better performance and the communication overhead is tolerable, increasing the size of the embedding vectors can be considered. However, if communication overhead is a significant limiting factor, a balance may need to be found between scale and performance to ensure that the communication budget is met and the performance requirements of the task are still met. A balance may need to be found between scale and performance to ensure that the communication budget is met and the performance requirements of the task are still met.

For convenience, we use $h(v_{k}^{j};\varphi)$ to denote the hypernetwork, and $\varphi$ to represent the hypernetwork parameters, with different $v_{k}^{j}$ inputs, the hypernetwork $h(v_{k}^{j};\varphi)$ can output the corresponding parameters. Each filter in the main network includes $N_{in} \times N_{out}$ kernels, and the size of each kernel is $f_{size} \times f_{size}$, which are mainly responsible for extracting various tiny features of the image for feature learning. The parameters of the filters use the $F_{j} \in \mathbb{R}^{N_{\text {in }} f_{\text {size }} \times N_{\text {out }} f_{\text {size }}}$ to indicate that, for each $F_j$, a hypernetwork is used to receive an embedding $v^j\in \mathbb{R}^{N_v}$ and to predict the outputs $F_j$, where $N_v$ is the size of the embedding vector. The hypernetwork we use is a two-layer linear network. The hypernetwork model flow is shown schematically in Fig.\ref{fig:Model}, where the embedding vector $v_k^j$ is linearly mapped as input to $N_{in}$ distinct matrices $W_{i}\in\mathbb{R}^{d\times N_{v}},i=1,...,N_{in}$ and the bias vector $B_{i}\in\mathbb{R}^{d},i=1,... ,N_{in}$, where $d$ is the size of the hidden layer in the hypernetwork. The input size of the second layer is $d$, using a matrix $W_{out}\in\mathbb{R}^{f_{size}\times{N_{out}}f_{size}\times d}$ and a bias matrix $B_{out}\in{\mathbb{R}^{f_{size}\times N_{out }f_{size}}}$ processed. Finally, the output is spliced into the parameters of a basic filter in a certain order.

An embedding vector $v_k^j$ represents a basic filter $F_j$ in the main network. The dimensions of the filters used for convolution operations in different convolution groups may be different, but are generally integer multiples of the basic filter. For example, use an embedding vector $v_k^j$ to represent the basic filter $F_j$ of $16\times 16$. If we want to represent $F_{32\times64}$, we can concatenate the hypernetwork outputs corresponding to multiple embedding vectors as parameters of a single filter. The following shows a $32 \times 64$ dimensional filter stitched together from 8 basic dimensions $16 \times 16$.
$$F_{32\times64}\quad=\quad\left(\begin{array}{cccc}F_1&F_2&F_3&F_4\\F_5&F_6&F_7&F_8\end{array}\right)$$

Generating all the parameters of the convolutional layer needed for a user $k$ requires all the embedding vector $\{v_k^j| j=1,2,... \}$, and for convenience, $v_k$ is used to represent all the feature embeddings of user $k$. Based on the above settings, the pFL objective of the HFN is adjusted to:
\begin{equation}
v^{*},\varphi^{*},w_{\beta}^{*}=\arg\min_{v,\varphi,w_{\beta} }\left[\frac{1}{\mathrm{K}}\sum_{\mathrm{k=1}}^{\mathrm{K}}\mathrm{f_k(h(v_{k};\varphi);w_{\beta_k}   )}\right] 
\end{equation}
where $v=\{v_1,\ldots,v_K,\}$ is the set of embedding vectors that can be learned by all users, and $w_{\beta}=\{w_{\beta_1},\ldots,w_{\beta_K}\}$ is the set of parameters of the personalization layer for all users. Instead of the traditional main network parameter $w$, what is aggregated in FL's server is the hypernetwork parameter $\varphi$, so the gradient $\bigtriangledown w$ of the main network needs to be further back-propagated to the hypernetwork, and the hypernetwork $\varphi$ is updated based on $\bigtriangledown w$ with $\varphi _{t+1} = \varphi_t - \lambda \bigtriangledown\varphi_t$ ($\lambda$ is the learning rate), thus the hypernetwork can perform end-to-end learning.

\subsection{Workflow}

\begin{algorithm}
  \caption{HyperFedNet}
  \label{alg:HFN}
  \DontPrintSemicolon
  \SetAlgoLined

  \textbf{Server executes:}

  \If{round = 1}{
    initialize $\varphi$\;
  }
    \For{round $t=1,2,\dots$}{
      $m\leftarrow\operatorname*{max}(C\cdot K,1)$\;
      $S_t\leftarrow(\text{random set of }m\text{ users})$\;
      \For{$k\in S_t$ \textbf{in parallel}}{
        $\varphi_{t+1}^k\leftarrow\text{UserUpdate}(k,\varphi_t)$\;
      }
      $\varphi_{t+1}\leftarrow\sum_{k=1}^K\frac{n_k}n\varphi_{t+1}^k$\;
    }
  
\SetKwProg{Fn}{Function}{:}{}
  \textbf{User executes:}
\SetKwProg{Fn}{Function}{:}{}

\If{$k$ is a new user}{initialize $w_{\beta}$ and $v$}

\Fn{UserUpdate{($k$, $\varphi$)}}
{
  $\mathcal{B} \leftarrow$ split $\mathcal{D}_k$ into batches of size $B$\;
  $w_{\theta} \leftarrow h(v, \varphi)$\;
  $w \leftarrow [w_{\theta}, w_{\beta}]$\;
  \For{local epoch $i$ from $1$ to $E$}{
    \For{batch $b \in \mathcal{B}$}{
      $w \leftarrow w - \eta \nabla f(w, b)$\;
      $\varphi \leftarrow \varphi - \eta \nabla f(w, b)$\;
      $v \leftarrow v - \eta \nabla f(w, b)$\;
    }
  }
  \Return $\varphi$\;
}

\end{algorithm}

We will explain the workflow of HFN. The workflow is divided into the server part (Line1-12) and the user part (Line13-28), as shown in Algorithm \ref{alg:HFN}.

\textbf{Server: }If it is in the first round, then initialize the parameter $\varphi$ (Line 2-4). Perform in each round (Line 5): select $m$ users with probability $C$ (Line 6), and put them into the set of users participating in training $S_t$ (Line 7), then send the hypernetwork parameter $\varphi_t$ to the selected users, who update the received parameter to the local hypernetwork, and then all the selected users perform local training in parallel (Lines 8-10). The updated parameters $\varphi$ are sent to the server after the training is completed (Line 28), and then the server aggregates hypernetwork parameters $\varphi_{t+1}$ for next round (Line 11). The process is repeated until the network converges or reaches the target round.

\textbf{User: }When a user first joins the training, user only need to initialize the classification layer, that is, the personalized parameters $w_{\beta}$, and the learnable embedding vector $v$ (Line 14-16). User update process (Line 17): First divide the local data set and use mini-batch for batch training (Line 18). Update the hypernetwork parameters $\varphi$ sent from the server to the local hypernetwork. Using the embedding vector $v$ as input, the output is generated according to the series rules to generate the main network basic layer parameters $w_{\theta}$ (Line 19), and then combined with the local personalized layer parameters $w_{\beta}$ to form a complete main network parameter $w$ (Line 20). Then perform local training, and perform gradient updates on $w,\varphi,v$ during the training process (Line 21-27). After completion, the locally updated hypernetwork parameters $\varphi$ will be sent back to the server, waiting for aggregation (Line 28).

The purpose of communicating the parameter $\varphi$ during the FL communication process is to allow the hypernetwork $h$ to be aggregated through the database training of different features, and to enhance its ability to generate quality parameters for the user's different filters to generate parameters $w_{\theta}$ suitable for extracting their own database features. After the FL training is completed, during the testing phase, the $w_{\theta}$ of the main network is still generated by the hypernetwork and deployed in the basic layer of the main network, at this time, the $w_{\beta}$ adapted is still the basic layer parameters from the previous round, so it is necessary to fine-tune a few local epochs by the user using the local data after the deployment of the basic layer parameters to make the personalization parameters $w_ {\beta}$ is better adapted to the newly generated parameters $w_{\theta}$.

\section{Experiments}
In this section,  we discuss the experiment setup, including the experimental environment and baseline algorithm. Firstly, we show how the hypernetwork generates parameters under similar data distributions. Subsequently, we highlight the advantages of the transmission hypernetwork over the transmission main network in terms of privacy preservation. We also report on the impact of varying embedding vector size on both communication volume and accuracy. Furthermore, we present the results of experiments conducted by the HFN on different real datasets. Finally, we show the effectiveness of combining HFN with other algorithms.

\subsection{Experiment Setup}

\begin{figure*}
\centering
\includegraphics[width=1.0\linewidth]{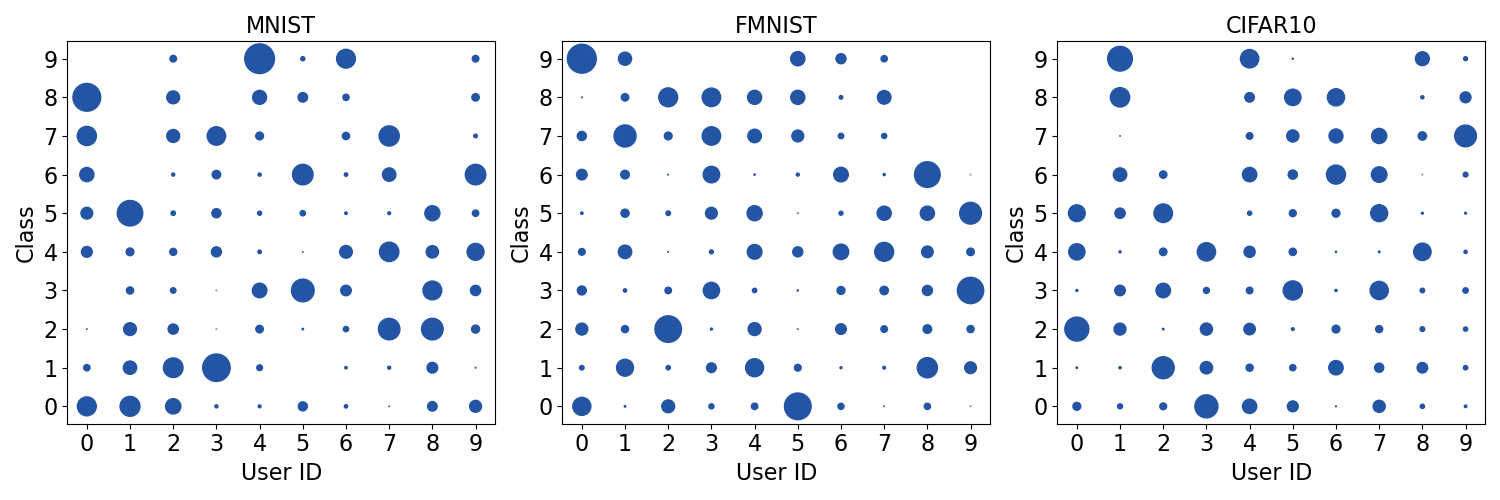}
\caption{Data distribution of $Dir(0.5)$.}
\label{fig:nonVis}
\end{figure*}

\begin{table*}[]
\centering
\caption{Convolutional Network Settings}
\label{tab:ResNetModel}
\begin{tabular}{c|cccc}
\hline
Group Name & MNIST                                                              & FMNIST                                                                    & CIFAR10                                                                   & CIFAR100                                                                    \\ \hline
conv1      & $\left[\begin{array}{c}3\times3,16\end{array}\right]$              & $\left[\begin{array}{c}3\times3,16\end{array}\right]$                     & $\left[\begin{array}{c}3\times3,16\end{array}\right]$                     & $\left[\begin{array}{c}3\times3,32\end{array}\right]$                       \\
conv2      & $\left[\begin{array}{c}3\times3,16\\3\times3,16\end{array}\right]$ & $\left[\begin{array}{c}3\times3,16\\3\times3,16\end{array}\right]\times6$ & $\left[\begin{array}{c}3\times3,16\\3\times3,16\end{array}\right]\times6$ & $\left[\begin{array}{c}3\times3,32\\3\times3,32\end{array}\right]\times6$   \\
conv3      & $\left[\begin{array}{c}3\times3,32\\3\times3,32\end{array}\right]$ & $\left[\begin{array}{c}3\times3,32\\3\times3,32\end{array}\right]\times6$ & $\left[\begin{array}{c}3\times3,32\\3\times3,32\end{array}\right]\times6$ & $\left[\begin{array}{c}3\times3,64\\3\times3,64\end{array}\right]\times6$   \\
conv4      & $\left[\begin{array}{c}3\times3,64\\3\times3,64\end{array}\right]$ & $\left[\begin{array}{c}3\times3,64\\3\times3,64\end{array}\right]\times6$ & $\left[\begin{array}{c}3\times3,64\\3\times3,64\end{array}\right]\times6$ & $\left[\begin{array}{c}3\times3,128\\3\times3,128\end{array}\right]\times6$ \\ \hline
\end{tabular}
\end{table*}

\textbf{Datasets and model: }We used four common datasets: MNIST\cite{MNIST-lecun1998gradient}, FMNIST\cite{FMNIST-xiao2017fashion}, CIFAR-10 and CIFAR-100\cite{CIFAR-krizhevsky2009learning}. We divided each dataset into 80\% training set and 20\% test set and divided them into 100 users. The test set and the training set on each user have the same data distribution, and the training set does not overlap with the test set data. The dirichlet distribution $Dir(0.5)$ is used to construct the non-iid case, in order to visualize the data heterogeneity among users, we plotted the heterogeneity of the first three datasets of 10 users at $Dir(0.5)$ distribution, as shown in Fig.\ref{fig:nonVis}, the presence or absence of dots represents whether there is such data, and the size of the dots represents the amount of such data.

The ResNet family \cite{he2016deep} is a representative architecture for CNN. Its network has many convolutional layers and therefore a large number of convolutional filters. We used it as the main network for experiments to validate the effect of the parameters generated by the hypernetwork. The architectures of the CNNs dealing with different datasets are shown in Table \ref{tab:ResNetModel}. These models are not the most advanced models for processing the dataset of this experiment, but are sufficient to demonstrate the purpose of the experiment, so we only need to focus on the relative performance. As an example, the ResNet architecture handles FMNIST and CIFAR10: the first set of convolutional blocks consists of one convolutional layer with an output channel of 16. The second set of convolutional blocks consists of 6 residual blocks, each with 2 convolutional layers, each with 16 output channels. The third and fourth sets of convolutional block composition are similar to the second set except that the output channels are 32 and 64, respectively. Convolution, batch normalization and ReLU activation are performed in the residual block in this order. The dimensions of the kernel are all $3 \times 3$. After convolution, an appropriate average pooling layer is used for dimensionality reduction. Finally, a fully connected classification layer is used. In the privacy experiment, we only want to verify the security effect and do not want to make the experiment very complicated, so we use the LeNet model\cite{MNIST-lecun1998gradient}.

\textbf{Baselines} 
In order to evaluate the performance of HFN, we compared it with the state-of-the-art FL algorithm. The unique hyperparameters of each algorithm in the experiment were tested by referring to the default configuration of the original paper. We implemented the following benchmark algorithms in the experiment: 

1) Centre: Centralize all data for training, imitating the traditional way of uploading data to the data center. This method cannot guarantee the security of data from all parties and has many shortcomings, but here we take it as the upper bound of FL accuracy. 

2) FedAvg\cite{fedavg-mcmahan2017communication}: This is the most important algorithm for FL, and it is effective in various scenarios. 

3) Local: There is no parameter exchange, but each user only uses his or her own data for training locally. 

4) FedBabu\cite{fedBabu-oh2021fedbabu}: The algorithm for updating and aggregating the model body needs fine-tuning after convergence. 

5) FedProx\cite{fedProx-li2020federated}: by adding proximal terms so that the user model does not deviate from the global model in order to deal with non-iid scenarios, we enumerated its unique hyperparameters $\mu=\{1, 0.1, 0.01\}$ and chose the optimal results for each experiment. 

6) FedGen\cite{FedGen-zhu2021data}: Improve FL accuracy by training a feature generator. 

7) FedDyn\cite{fedDyn-acar2021federated}: Propose a dynamic regularizer for each user in each round to promote consistent solutions for local and global users. 

8) pFedSim\cite{pFedSim-tan2023pfedsim}: Personalized algorithm based on model similarity. 

9) FedPer\cite{fedPer-arivazhagan2019federated}: A method of retaining the personalization layer locally, which can combat the adverse effects of statistical heterogeneity. 

10) FedBN\cite{fedbn-li2021fedbn}: Solve the problem of FL data heterogeneity by adding a batch normalization layer to the local model. 

11) FedRep\cite{fedrep-collins2021exploiting}: Train the classifier and feature extractor in sequence, and only aggregate the feature extractor. 

12) pFedla\cite{fedpFedLA-ma2022layer}: Deploys a hypernetwork on the server side for each user to give the user hierarchical aggregation weights.

13) FedFomo\cite{zhang2020personalized}: The optimal weighting of a customer is given to aggregate the model by calculating how much the customer can benefit from other customers' models. 

Since HFN and some of the algorithms require a fine-tuning step, to be fair, we add 4 local epochs of fine-tuning to all algorithms to ensure fairness and to evaluate their personalized accuracy.

\textbf{Settings}
The embedding vector size of HFN can be adjusted. In this general experiment, it is fixed, 64 when processing MNIST, and 128 for the rest. The experiment uses SGD as the optimizer, uses Nesterov Momentum, momentum is 0.9, weight decay is 5e-4, local epoch and fine-tuning epoch are 4, the batch size is set to 128, the total number of users is 100, the joining rate is 0.25, MNIST and FMNIST global communication round is 90, CIFAR10 and CIFAR100 are 150. We run multiple experiments on the learning rates of all the algorithms and select the best results in each experiment, including multi-step learning rates decaying from 0.1 and fixed learning rates of 0.1, 0.01, and 0.001.

\subsection{Hypernetwork for Parameter Generation}

\begin{figure*}
\centering
\includegraphics[width=1.0\linewidth]{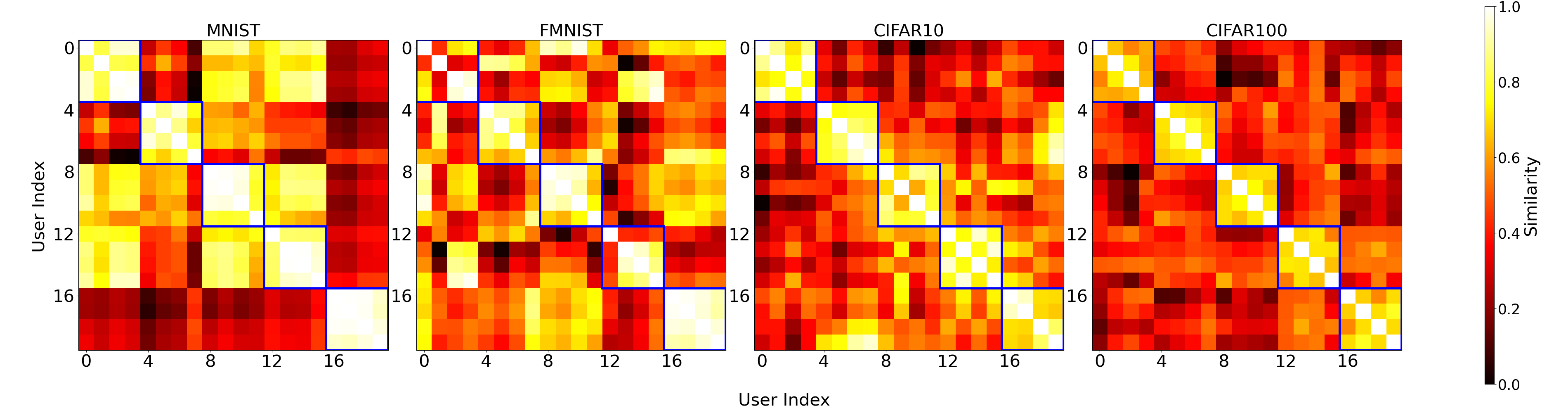}
\caption{Similarity of main network parameters generated by hypernetwork for different users.}
\label{fig:Heatmap}
\end{figure*}

In this experiment, there are a total of 20 users, each group of 4 users, with user numbers 0-3 as the first group, and so on, a total of 5 groups, and the users in the group have similar data distribution. Each user has 2 classes (MNIST, FMNIST, CIFAR10 dataset) or 10 classes (CIFAR100 dataset). We then evaluate the similarity of the generated parameters through cosine similarity. As shown in Fig.\ref{fig:Heatmap}, the images show the similarity of the hypernetwork generation parameters under the MNIST, FMNIST, CIFAR10 and CIFAR100 data sets. The horizontal and vertical coordinates in the figure are user numbers. The lighter the color of the grid, the higher the similarity of the user models corresponding to the x-axis and y-axis, and a similarity value of 1 means that the user models are identical. We use the blue border of $4\times4$ to highlight the similarity corresponding to the user models of the same group. It can be clearly seen from the figure that the color corresponding to the model within the group is significantly lighter, i.e., the similarity is higher, indicating that the user basic layer parameters within the group with similar data distribution have a higher similarity than the parameters in other groups. 

Users based on similar data distribution should have similarity in their embedding vector matrices. The parameters generated by the hypernetwork will be subject to similar conditions. This means that the hypernetwork will tend to generate parameters that adapt to similar data distributions, thus making the main network parameters generated by different users similar. However, it should be noted that even if users within a group have similar data distribution, the data content of each user is not completely consistent. The parameters generated by the hypernetwork for different users may vary to some extent. However, these differences should be within a certain range and maintain overall similarity.

This experiment shows that the hypernetwork we use can generate the basic layer parameters suitable for users with different data distributions by inputting the corresponding embedding vector to extract the unique features.

\subsection{Privacy Evaluation}

\begin{figure*}
\centering
\includegraphics[width=1.0\linewidth]{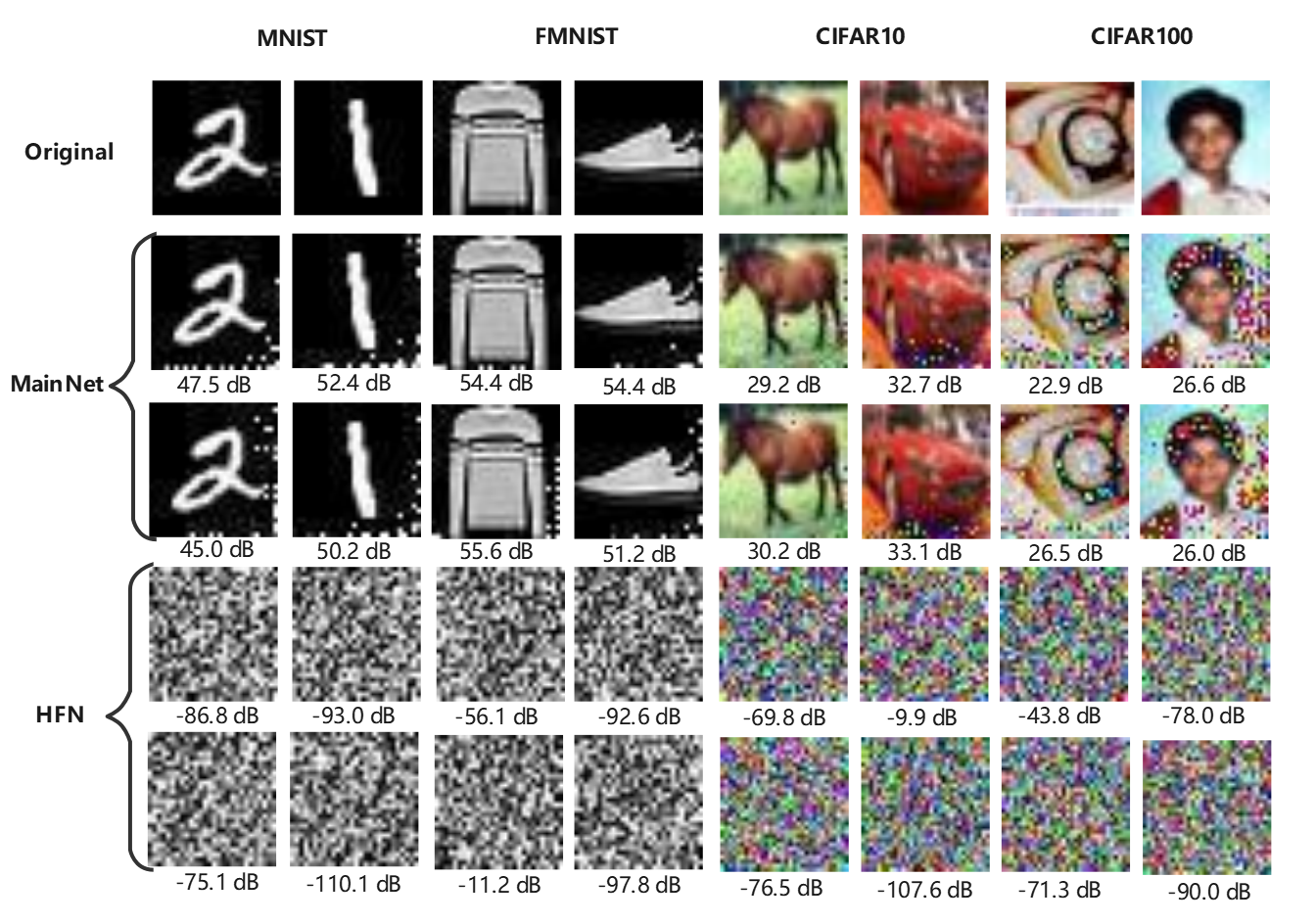}
\caption{Privacy attack results: two random images are listed for each dataset. The first row shows the data from the user's FL training. The second and third rows are the common FL methods used. The fourth and fifth rows are the HFN algorithm. The second and fourth rows are the results of the DLG attack. The third and fifth rows are the results of the iDLG attack. PSNR is reported under each recovered image.}
\label{fig:Privacy}
\end{figure*}

One of the advantages of FL is its ability to effectively protect user data privacy, however, a recent research paper \cite{yin2021see} pointed out that by obtaining the gradient information of the network during transmission, the intruder is able to infer the original data content, which raises concerns about the security of FL algorithms. In order to evaluate the security of FL algorithms, we have done on four different datasets attack experiments.

Peak signal-to-noise ratio (PSNR) is a measure of the reconstruction quality of an image compression signal. We calculate PSNR to represent the similarity between the original image and the reconstructed image. It is calculated as $PSNR=20\cdot\log_{10}\left(\frac{255}{\sqrt{MSE}}\right)$, where $MSE$ is the Mean Squared Error. The larger value of PSNR indicates that the attacked reconstructed image is more similar to the original image. According to Fig.\ref{fig:Privacy} shown, we can intuitively observe that the traditional FL method transmits the main network information with the risk of leaking data, which seriously threatens the basic principle of FL. On the contrary, the information transmitted by HFN algorithm is completely resistant to the attacks of DLG\cite{DLG-zhu2019deep} and iDLG\cite{iDLG-zhao2020idlg}. The HFN algorithm achieves true confidentiality and better protects the user's data privacy.

\subsection{The Affect of Embedding Vector Sizes on HFN}

\begin{figure}
\centering
\includegraphics[width=1.0\linewidth]{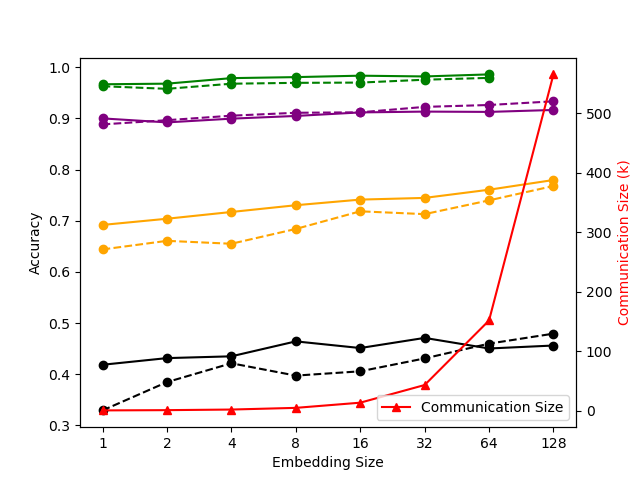}
\caption{The impact of different embedding sizes in HFN on communication size and accuracy.}
\label{fig:embeddingSize_ACC}
\end{figure}

HFN introduces a latent hyperparameter, the size of the Embedding Vector, which is related to the performance and communication overhead of FL. To investigate the relationship between them, we conducted experiments to explore the impact of different embedding sizes on accuracy and communication volume. As shown in Fig.\ref{fig:embeddingSize_ACC}, the red line represents the Cost per Round (CPR) for individual users, which corresponds to the right y-axis. CPR is composed of the parameter quantity transmitted by each user in one round, including both upload and download. It is important to note that the values represent the number of parameters transmitted in the network, not the actual network traffic. The other colors represent the accuracy achieved after convergence in FL, which corresponds to the left y-axis. The green line corresponds to MNIST, the purple line corresponds to FMNIST, the orange line corresponds to CIFAR-10, and the black line corresponds to CIFAR100. Comparing the iid and non-iid (solid and dashed lines), it can be observed that in the non-iid setting, as the embedding vector size increases, the improvement in accuracy is more significant compared to the iid case. This is because in the non-iid scenario, the variations in data distributions among different users require the convolutional layers of different users to extract more diverse feature information. This necessitates a larger embedding vector for assistance, resulting in a notable increase in accuracy as the embedding vector size increases.

Regarding communication overhead, the communication cost of HFN increases with the increase in embedding size. The communication overhead of most FL algorithms is similar to FedAvg. We calculated the CPR of FedAvg for individual users for the MNIST, FMNIST, CIFAR10, and CIFAR100 datasets, which are 155K, 1127K, 1128K, and 4519K, respectively. Comparing the different embedding sizes of HFN in Fig.\ref{fig:embeddingSize_ACC}, the communication overhead is significantly higher. However, when employing an embedding vector size of 1, the CPR of HFN on the four datasets is only 640. Therefore, the HFN algorithm has an advantage in terms of communication overhead. This graph allows us to flexibly balance accuracy and communication overhead based on practical requirements, enabling the selection of an appropriate embedding vector size.

\subsection{Performance Evaluation}

\begin{table*}[]
\centering
\caption{The accuracy of the algorithms under four databases}
\label{tab:acc}
\begin{tabular}{ccccccccc}
\hline
                                 & \multicolumn{2}{c}{MNIST}                                            & \multicolumn{2}{c}{FMNIST}                                           & \multicolumn{2}{c}{CIFAR10}                                          & \multicolumn{2}{c}{CIFAR100}                   \\ \cline{2-9} 
aggregation method                     & non-iid                & \multicolumn{1}{c|}{iid}                    & non-iid                & \multicolumn{1}{c|}{iid}                    & non-iid                & \multicolumn{1}{c|}{iid}                    & non-iid                & iid                   \\ \hline
\multicolumn{1}{c|}{Centre}     & \multicolumn{2}{c|}{99.49\%}                                         & \multicolumn{2}{c|}{93.60\%}                                          & \multicolumn{2}{c|}{84.78\%}                                         & \multicolumn{2}{c}{54.78\%}                    \\
\multicolumn{1}{c|}{FedAvg}      & 98.67\%                & \multicolumn{1}{c|}{98.96\%}                & 92.40\%                & \multicolumn{1}{c|}{90.12\%}                & 69.95\%                & \multicolumn{1}{c|}{62.12\%}                & 33.85\%                & 33.67\%               \\
\multicolumn{1}{c|}{Local}       & 94.82\%                & \multicolumn{1}{c|}{93.35\%}                & 86.94\%                & \multicolumn{1}{c|}{76.74\%}                & 57.05\%                & \multicolumn{1}{c|}{30.45\%}                & 20.15\%                & 6.98\%                \\
\multicolumn{1}{c|}{FedBabu}     & 98.94\%                & \multicolumn{1}{c|}{98.81\%}                & {\ul \textbf{94.12\%}} & \multicolumn{1}{c|}{91.04\%}                & 73.34\%                & \multicolumn{1}{c|}{64.39\%}                & 43.79\%                & 36.67\%               \\
\multicolumn{1}{c|}{FedProx}     & 98.85\%                & \multicolumn{1}{c|}{{\ul \textbf{99.04\%}}} & 93.10\%                & \multicolumn{1}{c|}{90.39\%}                & 69.89\%                & \multicolumn{1}{c|}{60.88\%}                & 31.14\%                & 34.17\%               \\
\multicolumn{1}{c|}{FedGen}      & 97.63\%                & \multicolumn{1}{c|}{98.18\%}                & 91.78\%                & \multicolumn{1}{c|}{89.36\%}                & 67.06\%                & \multicolumn{1}{c|}{62.75\%}                & 26.32\%                & 16.06\%               \\
\multicolumn{1}{c|}{FedDyn}      & 98.73\%                & \multicolumn{1}{c|}{99.01\%}                & 93.20\%                & \multicolumn{1}{c|}{90.64\%}                & 67.76\%                & \multicolumn{1}{c|}{61.60\%}                & 29.23\%                & 34.37\%               \\
\multicolumn{1}{c|}{pFedSim}     & 98.92\%                & \multicolumn{1}{c|}{99.01\%}                & 92.41\%                & \multicolumn{1}{c|}{90.47\%}                & 70.45\%                & \multicolumn{1}{c|}{65.66\%}                & 43.74\%                & 37.30\%                \\
\multicolumn{1}{c|}{FedPer}      & 99.05\%                & \multicolumn{1}{c|}{98.84\%}                & 92.36\%                & \multicolumn{1}{c|}{89.59\%}                & 72.67\%                & \multicolumn{1}{c|}{62.63\%}                & 31.15\%                & 17.10\%                \\
\multicolumn{1}{c|}{FedBN}       & {\ul \textbf{99.14\%}} & \multicolumn{1}{c|}{98.94\%}                & 91.64\%                & \multicolumn{1}{c|}{89.74\%}                & 69.16\%                & \multicolumn{1}{c|}{33.77\%}                & 35.39\%                & 32.08\%               \\
\multicolumn{1}{c|}{FedRep}      & 98.10\%                & \multicolumn{1}{c|}{98.01\%}                & 92.36\%                & \multicolumn{1}{c|}{87.18\%}                & 75.44\%                & \multicolumn{1}{c|}{65.59\%}                & 28.77\%                & 12.38\%               \\
\multicolumn{1}{c|}{pFedla}      & 98.40\%                & \multicolumn{1}{c|}{98.49\%}                & 92.78\%                & \multicolumn{1}{c|}{88.64\%}                & 66.18\%                & \multicolumn{1}{c|}{56.98\%}                & 29.54\%                & 26.17\%               \\
\multicolumn{1}{c|}{FedFomo}     & 93.95\%                & \multicolumn{1}{c|}{92.02\%}                & 86.09\%                & \multicolumn{1}{c|}{76.06\%}                & 56.29\%                & \multicolumn{1}{c|}{24.85\%}                & 17.71\%                & 12.02\%               \\ \hline
\multicolumn{1}{c|}{HFN(Ours)} & 97.91\%                & \multicolumn{1}{c|}{98.59\%}                & 93.33\%                & \multicolumn{1}{c|}{{\ul \textbf{91.64\%}}} & {\ul \textbf{76.78\%}} & \multicolumn{1}{c|}{{\ul \textbf{77.95\%}}} & {\ul \textbf{47.92\%}} & {\ul \textbf{45.60\%}} \\ \hline
\end{tabular}
\end{table*}

\begin{table*}[]
\centering
\caption{Communication cost for a single user at each level of the different algorithms when the dataset is CIFAR10-iid}
\label{tab:comm}
\begin{tabular}{ccccccccc}
\hline
 & CPR   & 10\%              & 20\%                & 30\%                & 40\%                 & 50\%                 & 60\%                 & 70\%                \\ \hline
HFN\_128   & 0.57M & 0.00M $(0\times)$ & 1.13M $(2\times)$   & 7.92M $(14\times)$  & 10.75M $(19\times)$  & 15.84M $(28\times)$  & 23.19M $(41\times)$  & 35.63M $(63\times)$ \\
FedAvg      & 1.13M & 1.13M $(1\times)$ & 5.64M $(5\times)$   & 15.79M $(14\times)$ & 46.24M $(41\times)$  & 46.24M $(41\times)$  & 91.36M $(81\times)$  & \textbackslash{}    \\
FedBabu     & 1.13M & 1.13M $(1\times)$ & 2.25M $(2\times)$   & 7.89M $(7\times)$   & 25.91M $(23\times)$  & 46.19M $(41\times)$  & 46.19M $(41\times)$  & \textbackslash{}    \\
FedProx     & 1.13M & 1.13M $(1\times)$ & 14.66M $(13\times)$ & 46.24M $(41\times)$ & 46.24M $(41\times)$  & 72.18M $(64\times)$  & 92.48M $(82\times)$  & \textbackslash{}    \\
FedGen      & 1.13M & 1.13M $(1\times)$ & 28.29M $(25\times)$ & 39.60M $(35\times)$ & 46.39M $(41\times)$  & 89.38M $(79\times)$  & 91.65M $(81\times)$  & \textbackslash{}    \\
FedDyn      & 1.13M & 1.13M $(1\times)$ & 3.38M $(3\times)$   & 14.66M $(13\times)$ & 34.96M $(31\times)$  & 46.24M $(41\times)$  & 91.36M $(81\times)$  & \textbackslash{}    \\
pFedSim     & 1.13M & 1.13M $(1\times)$ & 5.64M $(5\times)$   & 15.79M $(14\times)$ & 46.24M $(41\times)$  & 46.24M $(41\times)$  & 109.40M $(97\times)$ & \textbackslash{}    \\
FedPer      & 1.13M & 1.13M $(1\times)$ & 6.76M $(6\times)$   & 16.90M $(15\times)$ & 28.16M $(25\times)$  & 46.19M $(41\times)$  & 91.25M $(81\times)$  & \textbackslash{}    \\
FedBN       & 1.13M & 1.13M $(1\times)$ & 4.51M $(4\times)$   & 15.79M $(14\times)$ & 46.24M $(41\times)$  & 46.24M $(41\times)$  & 91.36M $(81\times)$  & \textbackslash{}    \\
FedRep      & 1.13M & 0.00M $(0\times)$ & 9.01M $(8\times)$   & 15.77M $(14\times)$ & 40.56M $(36\times)$  & 78.86M $(70\times)$  & \textbackslash{}     & \textbackslash{}    \\
pFedla      & 1.13M & 0.00M $(0\times)$ & 16.92M $(15\times)$ & 27.07M $(24\times)$ & 46.24M $(41\times)$  & 91.36M $(81\times)$  & \textbackslash{}     & \textbackslash{}    \\
FedFomo     & 3.95M & 0.00M $(0\times)$ & 15.79M $(4\times)$  & 35.53M $(9\times)$  & 221.06M $(56\times)$ & 307.91M $(78\times)$ & \textbackslash{}     & \textbackslash{}    \\ \hline
\end{tabular}
\end{table*}

\begin{table*}[]
        \setlength{\tabcolsep}{2pt}
        \centering
        \caption{The effect of HFN combined with other algorithms}
        \label{table:tab_HFN_others}
	\scalebox{1.0}{
	\centering
	\begin{tabular}{ccccc|cccc|cccc|cccc}
		
		\hline
		& \multicolumn{4}{c|}{MNIST}                                        & \multicolumn{4}{c|}{FMNIST}                                       & \multicolumn{4}{c|}{CIFAR10}                                       & \multicolumn{4}{c}{CIFAR100}                                      \\ \cline{2-17} 
		& \multicolumn{2}{c|}{non-iid}           & \multicolumn{2}{c|}{iid} & \multicolumn{2}{c|}{non-iid}           & \multicolumn{2}{c|}{iid} & \multicolumn{2}{c|}{non-iid}            & \multicolumn{2}{c|}{iid} & \multicolumn{2}{c|}{non-iid}            & \multicolumn{2}{c}{iid} \\
		& Acc     & \multicolumn{1}{c|}{Comm}    & Acc         & Comm       & Acc     & \multicolumn{1}{c|}{Comm}    & Acc         & Comm       & Acc     & \multicolumn{1}{c|}{Comm}     & Acc         & Comm       & Acc     & \multicolumn{1}{c|}{Comm}     & Acc        & Comm       \\ \hline
		\multicolumn{1}{c|}{HFN+FedAvg}  & -1.02\% & \multicolumn{1}{c|}{15.15\%} & -0.75\%     & 13.46\%    & 0.98\%  & \multicolumn{1}{c|}{50.14\%} & 1.41\%      & 61.29\%    & 13.38\% & \multicolumn{1}{c|}{51.95\%}  & 15.33\%     & 74.60\%    & 13.13\% & \multicolumn{1}{c|}{78.80\%}  & 13.63\%    & 48.98\%    \\
		\multicolumn{1}{c|}{HFN+FedProx} & -0.90\% & \multicolumn{1}{c|}{18.27\%} & -0.55\%     & 19.41\%    & 0.25\%  & \multicolumn{1}{c|}{54.32\%} & 1.02\%      & 71.31\%    & 8.14\%  & \multicolumn{1}{c|}{78.88\%}  & 16.98\%     & 50.55\%    & 16.19\% & \multicolumn{1}{c|}{164.75\%} & 12.51\%    & 48.98\%    \\
		\multicolumn{1}{c|}{HFN+FedDyn}  & -0.82\% & \multicolumn{1}{c|}{22.08\%} & -0.55\%     & 18.55\%    & -0.34\% & \multicolumn{1}{c|}{60.64\%} & 1.22\%      & 50.14\%    & 9.30\%  & \multicolumn{1}{c|}{140.62\%} & 16.38\%     & 70.93\%    & 18.18\% & \multicolumn{1}{c|}{172.58\%} & 11.67\%    & 51.34\%    \\
		\multicolumn{1}{l|}{HFN+FedBN}   & -0.86\% & \multicolumn{1}{c|}{18.40\%} & -1.29\%     & 12.84\%    & -0.89\% & \multicolumn{1}{c|}{71.45\%} & 1.23\%      & 50.14\%    & 0.16\%  & \multicolumn{1}{c|}{51.07\%}  & 36.67\%     & 94.83\%    & 4.58\%  & \multicolumn{1}{c|}{158.78\%} & 5.29\%     & 49.55\%    \\
		\multicolumn{1}{l|}{HFN+pFedla}  & -0.38\% & \multicolumn{1}{c|}{12.00\%} & -0.21\%     & 15.86\%    & 0.22\%  & \multicolumn{1}{c|}{54.32\%} & 3.13\%      & 71.31\%    & 10.92\% & \multicolumn{1}{c|}{119.78\%} & 20.17\%     & 50.14\%    & 16.52\% & \multicolumn{1}{c|}{152.66\%} & 19.51\%    & 34.52\%    \\
		\multicolumn{1}{l|}{HFN+FedFomo} & 3.95\%  & \multicolumn{1}{c|}{24.43\%} & 6.47\%      & 32.10\%    & 6.49\%  & \multicolumn{1}{c|}{21.88\%} & 15.53\%     & 35.00\%    & 17.19\% & \multicolumn{1}{c|}{71.62\%}  & 50.23\%     & 33.12\%    & 20.87\% & \multicolumn{1}{c|}{34.04\%}  & 28.54\%    & 11.40\%    \\ \hline
	\end{tabular}
}
\end{table*}

The personalization accuracies under the same communication round limit are shown in Table \ref{tab:acc}. In most cases, the HFN algorithm demonstrates good performance, especially when the dataset is complex. When HFN comes to process the simpler MNIST dataset, average performance may be observed. This can be attributed to the relative simplicity of the MNIST dataset and the simpler model structure used. In the MNIST dataset, the image has a lower image resolution and the digit patterns are relatively simple, so a simple model can be used to extract enough features for classification. In this case, the introduction of a hypernetwork may increase the complexity of the model and make the task relatively more complex, resulting in less remarkable accuracy. It is worth noting that hypernetworks were originally designed to handle more complex tasks and models. It is when applied to more complex datasets (CIFAR series) and models that the benefits of HFN become more significant. This is because the choice of model parameters can be more critical in complex tasks, and the hypernetwork can provide more appropriate parameters by adaptively generating model parameters that produce higher accuracy.

Communication cost has always been a problem in FL, but the HFN algorithm can solve this problem effectively. In Table \ref{tab:comm}, we show the communication cost of different algorithms during the training process when users are trained with the CIFAR-10 dataset (iid). Also, we list the communication costs incurred by each algorithm in reaching a certain accuracy threshold. The number in parentheses indicates the rounds in which the algorithm reaches that accuracy threshold, by which we can roughly determine the convergence speed of each algorithm. It can be seen that HFN not only has a significant advantage in convergence speed, but also has an unrivaled advantage in communication cost. It is worth noting that the advantage of HFN in reducing communication costs is due to its ability to generate model parameters locally without the need to transmit large amounts of raw data. This local generation of parameters reduces the amount of data transfer between users and improves the efficiency of FL. Therefore, HFN can better solve the problem of communication overhead in FL by reducing the amount of data transfer between users while producing better accuracy.

\subsection{HFN with Other FL Algorithm}

Another major advantage of HFN is that it can be easily combined with existing algorithms, by which the advantages of HFN can be attached to existing algorithms and benefit from other algorithms. We simply experimented by fusing HFN with other partially baseline algorithms, and obtained the Table \ref{table:tab_HFN_others}. Where ACC indicates how much the algorithm has improved on the original accuracy after fusion, for example, the accuracy of FedAvg at CIFAR10 non-iid is 69.95\%, and HFN+FedAvg is 83.33\%, then the corresponding value of ACC in the table should be recorded as 13.38\%. We also counted the corresponding communication consumption of the algorithms from the beginning of training, all the way to the final accuracy, compared with the fused algorithms. The percentage in Comm means the percentage of the total communication consumption of the new algorithms after fusing the HFN algorithms to the original total consumption. It can be noticed that the communication efficiency of all the baseline algorithms is greatly improved by combining HFN. In some cases, the total amount of communication exceeds 100$\%$, which is reasonable due to the fact that the convergence accuracy has been improved more than the original and more communication rounds are needed to achieve higher accuracy.

\section{Conclusion}
In this work, we explore and utilize the potential of hypernetwork in FL. Compared with the previous traditional architecture, using the hypernetwork instead of the main network for communication learning greatly saves the communication cost, improves safety and the powerful learning ability of the hypernetwork also improves the accuracy of the main network. We verified the performance of traditional algorithms and HFN under different datasets with different distributions through a large number of experiments, and then did many targeted experiments on HFN and fused this novel approach into the traditional FL method to achieve better results.

\bibliographystyle{unsrt}
\bibliography{sample}
 
\vspace{11pt}

\vspace{-33pt}

\vspace{11pt}

\vfill

\end{document}